\documentclass[a4paper,fleqn,usenatbib]{mnras}
\usepackage[T1]{fontenc}
\usepackage{ae,aecompl}
\usepackage{graphicx}
\usepackage{tabularx}
\usepackage{txfonts}
\usepackage{amssymb}	

\title[Hadronic $\gamma$-rays from extreme TeV blazars]{Observable spectral and angular distributions of $\gamma$-rays\\ from extragalactic ultrahigh energy cosmic ray accelerators:\\ the case of extreme TeV blazars}
\author[E.V. Khalikov \& T.A. Dzhatdoev]{
Emil V. Khalikov,$^{1}$\thanks{E-mail: nanti93@mail.ru}
Timur A. Dzhatdoev$^{1,2}$\thanks{E-mail: timur1606@gmail.com}
\\
$^{1}$Federal State Budget Educational Institution of Higher Education, M.V. Lomonosov Moscow State University, \\Skobeltsyn Institute of Nuclear Physics (SINP MSU), 1(2), Leninskie gory, GSP-1, 119991 Moscow, Russia\\
$^{2}$Institute for Cosmic Ray Research, University of Tokyo, 5-1-5 Kashiwanoha, 277-8568 Kashiwa, Japan\\
}

\date{Accepted XXX. Received YYY; in original form ZZZ}

\pubyear{2020}

\begin{document}
\label{firstpage}
\pagerange{\pageref{firstpage}--\pageref{lastpage}}
\maketitle

\begin{abstract}
Ultrahigh energy protons and nuclei from extragalactic cosmic ray sources initiate intergalactic electromagnetic cascades, resulting in observable fluxes of $\gamma$-rays in the GeV-TeV energy domain. The total spectrum of such cascade $\gamma$-rays of hadronic nature is significantly harder than the one usually expected from blazars. The spectra of some sources known as ``extreme TeV blazars'' could be well-described by this ``intergalactic hadronic cascade model'' (IHCM). We calculate the shape of the observable point-like spectrum, as well as the observable angular distibution of $\gamma$-rays, for the first time taking into account the effect of primary proton deflection in filaments and galaxy clusters of the extragalactic magnetic field assuming the model of Dolag et al. (2005). We present estimates of the width of the observable $\gamma$-ray angular distribution derived from simple geometrical considerations. We also employ a hybrid code to compute the observable spectral and angular distributions of $\gamma$-rays. The observable point-like spectrum at multi-TeV energies is much softer than the one averaged over all values of the observable angle. The presence of a high-energy cutoff in the observable spectra of extreme TeV blazars in the framework of the IHCM could significantly facilitate future searches of new physics processes that enhance the apparent $\gamma$-ray transparency of the Universe (for instance, $\gamma \rightarrow ALP$ oscillations). The width of the observable angular distribution is greater than or comparable to the extent of the point spread
function of next-generation $\gamma$-ray telescopes.
\end{abstract}


\section{Introduction}

Observations of ultrahigh energy cosmic rays (UHECR) indicate that these particles are mostly extragalactic, at least when the primary particle energy $E_{0}$ exceeds 10$^{19}$ eV = 10~EeV \citep{PierreAuger2017}\footnote {see, however, \citet{Gruzinov2018} and \citet{Strong2018}}. Active galactic nuclei (AGN) and, in particular, blazars (AGN with the jets presumably pointed towards the observer) are among the most well-motivated sources of UHECR \citep{Hillas1984,Biermann1987,Rachen1993,Dermer2010}.

Ultrahigh energy protons and nuclei from extragalactic sources interact with diffuse photon backgrounds by means of photohadronic processes and electron-positron pair production \citep{Greisen1966,Zatsepin1966,Berezinsky1988,Berezinsky2006}\footnote {in what follows we will mainly consider the case of primary protons}. Secondary $\gamma$-rays, electrons and positrons (hereafter ``electrons'' for simplicity) develop intergalactic electromagnetic (EM) cascades \citep{Berezinsky1975,Protheroe1986,Protheroe1993,Aharonian1994,Berezinsky2016} via the pair production (PP) process ($\gamma\gamma \rightarrow e^{+}e^{-}$) and the inverse Compton (IC) scattering ($e^{+}\gamma \rightarrow e^{+'}\gamma^{'}$ or $e^{-}\gamma \rightarrow e^{-'}\gamma^{'}$). Cascade $\gamma$-rays that typically have the energy much lower than that of the primary protons may be observed with space $\gamma$-ray telescopes such as {\it \mbox{Fermi-LAT}} \citep{Atwood2009}, {\it \mbox{AGILE}} \citep{Tavani2009,Bulgarelli2019} or projected space observatories {\it \mbox{MAST}} \citep{Dzhatdoev2019a}, {\it \mbox{GAMMA-400}} \citep{Galper2013}, {\it \mbox{e-ASTROGAM}} \citep{DeAngelis2018}, {\it \mbox{AMEGO}} \citep{McEnery2019}, {\it \mbox{AdEPT}} \citep{Hunter2014}, {\it \mbox{HARPO}} \citep{Gros2018}, or with ground-based $\gamma$-ray detectors such as \textit{\mbox{H.E.S.S.}} \citep{Hinton2004,Bonnefoy2018}, \textit{\mbox{MAGIC}} \citep{Aleksic2016a,Aleksic2016b}, \textit{\mbox{VERITAS}} \citep{Krennrich2004,Park2016}, \textit{\mbox{HAWC}} \citep{Abeysekara2017}, Cherenkov Telescope Array (\textit{\mbox{CTA}}) \citep{Actis2011,Acharya2013,CTA2018}, \textit{\mbox{ASTRI}} \citep{Lombardi2019}, \textit{\mbox{LHAASO}} \citep{Cui2014,Tian2018,LHAASO-arxiv-2019}.

The observable angular distribution of these cascade $\gamma$-rays from a point-like source strongly depends on the strength and structure of the extragalactic magnetic field (EGMF). In particular, if the deflection angles of primary protons and cascade electrons are small enough, the observational appearance of the source is similar to a slightly extended $\gamma$-ray pattern. The idea that cascade $\gamma$-rays from primary protons accelerated in extragalactic sources may contribute to point-like images of these sources was considered in \citet{Waxman1996} and \citet{Uryson1998}, thus laying the foundation for the ``intergalactic hadronic cascade model'' (IHCM). \citet{Waxman1996} mainly concentrated on temporal properties of the $\gamma$-ray signal and a general discussion of the IHCM, while \citet{Uryson1998} demonstrated that a part of the observable $\gamma$-rays may have a multi-TeV energy. The appearance of these very energetic $\gamma$-rays is due to the fact that a part of the interactions of the primary protons occurs relatively near to the observer, and thus the cascade $\gamma$-rays experience a lesser degree of absorption on diffuse extragalactic photon backgrounds than the primary $\gamma$-rays from the same source. $\gamma$-ray spectra of EM cascades initiated by UHECR injected at cosmological distances were calculated in \citep{Kalashev2001,Kalashev2009}.

However, in $\gamma$-ray astronomy it is still customary to neglect any effects from intergalactic EM cascades initiated by either $\gamma$-rays or protons, assuming the so-called ``absorption-only model'' (AOM) that accounts for only PP and adiabatic losses. In the framework of the AOM one can reconstruct the shape of the intrinsic spectrum of the source by compensating for the attenuation of the primary $\gamma$-ray flux and redshift. Observations made with imaging atmospheric Cherenkov telescopes (IACTs) revealed that some blazars have such reconstructed intrinsic spectral energy distributions (SEDs) peaking at $E > 1$ TeV (e.g. \citet{Aharonian2006,Aliu2014}). In what follows we call these AGNs ``extreme TeV blazars'' (ETBs).

ETBs are defined solely by their $\gamma$-ray properties, but they have much in common with extreme highly peaked BL Lac objects (EHBLs) \citep{Costamante2001,Bonnoli2015,Costamante2018}, which may be characterised by their broadband (in particular, X-ray) properties. Indeed, some blazars such as 1ES 0229+200 \citep{Aharonian2007a,Aliu2014}, 1ES 1101-232 \citep{Aharonian2006,Aharonian2007b} and 1ES 0347-121 \citep{Aharonian2007c} may be classified as both ETBs (see \citet{Dzhatdoev2017}, Fig. 12, 14, 17) and EHBLs. Compared to well-known nearby blazars such as Mkn 501 and Mkn 421, ETBs, as a rule, reveal weak and slow variability in the high energy (HE, $E > 100$ MeV) and the very high energy (VHE, $E > 100$ GeV) spectral bands.

These peculiar properties of ETBs allow for a possibility that a significant part of observable $\gamma$-rays were in fact produced not inside the source, but as a result of EM cascade development in the intergalactic medium. Therefore, the IHCM experienced a very high level of popularity when applied to ETBs and EHBLs \citep{Essey2010,Essey2011,Murase2012,Zheng2015,Archer2018,Acciari2019a,Acciari2019b,Das2019}; for a review see \citep{Biteau2020}.

Even before the discovery of ETBs in the VHE energy range, first sophisticated models of the EGMF in the large-scale structure (LSS) were developed \citep{Dolag2005,Sigl2004,Das2008,Vazza2017,AlvesBatista2017,Hackstein2018,Hackstein2019}. Most of these models predict an appreciable ($\sim 0.1-1^{\circ}$) deflection for UHE protons after propagating $\sim 100$ Mpc, and the EGMF appears to be strongly inhomogeneous \citep{Ryu1998,MedinaTanco1998}, with relatively strong magnetic fields in galaxy clusters ($B \sim 1-10$ $\mu$G), $B \sim 1-100$ nG in filaments, and comparatively weak ($B < 1$ nG) EGMF in voids. There is mounting evidence for the existence of magnetic fields in extragalactic filaments. In particular, \citet{Govoni2019} report the detection of synchrotron emission from a filament connecting two galaxy clusters (see also \citet{Vacca2018}, where the detection of another filament of the cosmic web was reported).

In the present paper, for the first time, we show how primary proton deflections in extragalactic magnetic field described by large scale structure formation models (in particular, strongly inhomogeneous EGMF) modify the observable spectral and angular distributions of $\gamma$-rays. We demonstrate that this effect broadens the observable angular distribution significantly and leads to an effective cutoff in the observable point-like spectrum\footnote {that is, the spectrum inside the point spread function (PSF) of the observing instrument} (in what follows called simply ``observable spectrum''). Some preliminary estimates justifying these conclusions were published in our recent work \citet{Dzhatdoev2019b} as conference contribution. The deflection of cascade electrons in weaker magnetic fields of LSS voids may contribute to the broadening of the observable angular distribution, leading to an even more dramatic effect.

The present work was in part motivated by the on-going search for oscillations of $\gamma$-rays to axion-like particles ($\gamma \rightarrow ALP$) using blazar spectra \citep{Horns2012,Rubtsov2014,Dzhatdoev2015,Korochkin2019a}. $\gamma$-rays may convert to ALPs \citep{Raffelt1988,DeAngelis2007,Kartavtsev2017,Montanino2017,Galanti2018,Galanti2019a} in magnetic fields of the source or in the EGMF clusters or filaments; the produced ALPs then may convert back to observable $\gamma$-rays. A part of the path of these observable $\gamma$-rays is, thus, traversed by ALPs that do not experience absorption on extragalactic background light (EBL) photons; therefore, the optical depth $\tau$ for $\gamma$-rays is effectively reduced and the Universe appears more transparent in the presence of the $\gamma \rightarrow ALP$ oscillation process.

The discovery of ALPs by means of this effect would require a detailed estimation of astrophysical backgrounds for the $\gamma \rightarrow ALP$ process and then a suppression of these backgrounds. Cascade $\gamma$-rays from primary proton interactions in fact represent the most dangerous source of the background for these $\gamma \rightarrow ALP$ oscillation searches \citep{Baklagin2018}. The presence of an effective cutoff in the observable spectrum with respect to the angle-averaged spectrum in the framework of the IHCM allows to greatly reduce the background for $\gamma \rightarrow ALP$ oscillation search in blazar spectra at high values of the $\gamma\gamma$ opacity.

This paper is organized as follows. In Sect.~\ref{sec:estimates} we present some basic estimates of the angular broadening and spectral cutoff effects, devoid of any unnecessary computational details. In Sect.~\ref{sec:simulations} we describe simulations of proton deflection and calculation of the observable spectrum by means of a hybrid method introduced by us in \citet{Dzhatdoev2017} (hereafter denoted as D17). We present the main results of this paper in Sect.~\ref{sec:results} and observational prospects --- in Sect.~\ref{sec:prospects}. We briefly discuss these results and conclude in Sect.~\ref{sec:conclusions}.

Throughout this work we assume the following values of cosmological parameters: $H_{0} = 67.8$ km s$^{-1}$ Mpc$^{-1}$, $\Omega_{m} = 0.308$, $\Omega_{\Lambda} \approx 1-\Omega_{m}$ \citep{Planck2016}. Somewhat different results for these parameters were presented in \citet{Planck2018}: $H_{0} = 67.4$ km s$^{-1}$ Mpc$^{-1}$ and $\Omega_{m} = 0.315$; these updated values, however, would not influence our results and conclusions significantly.

\section{Estimates} \label{sec:estimates}

Primary protons deflect to appreciable angles while they propagate over a large distance $L_{S} > 100$ Mpc from the source to the observer. According to the EGMF model of  \citet{Dolag2005} (hereafter denoted as D05), a half of protons with the energy $E_{p} = 40$ EeV deflect to the angle $\delta > 0.7^{\circ}$ for $L_{S} = 500$ Mpc (see Fig. 15 in the aforementioned paper). For larger distances, neglecting proton energy losses, we estimate 
\begin{equation}
 0.7^{\circ}\sqrt{\frac{L_{S}}{500 \: Mpc}}\frac{40 \: EeV}{E_{p}} < \delta < 0.7^{\circ}\frac{L_{S}}{500 \: Mpc}\frac{40 \: EeV}{E_{p}}; \label{eq1}
\end{equation}
in particular, for $E_{p} = 30$ EeV and $z_{s} = 0.186$ $1.2^{\circ} < \delta < 1.5^{\circ}$ (throughout this section we conservatively assume $\delta = 1^{\circ}$).\footnote{the account of energy losses would increase the value of $\delta$}
The total value of  $\delta$ may be dominated by one or two of the most prominent magnetic structures, or may accumulate over many less significant ones. Below we estimate the width of the observable $\gamma$-ray angular distribution separately for these cases.

In the present work we concentrate on the impact that primary proton deflections have on the observable spectral and angular distributions of $\gamma$-rays. Cascade electrons also experience deflections on magnetic fields of filaments and voids. The electrons that were produced in filaments are typically almost completely isotropized (their deflection angles $\delta_{e} \gg 1$ rad), and thus the secondary (cascade) $\gamma$-rays form a quasi-isotropic cloud (the so-called ``pair halo'', PH), as discussed in \citet{Aharonian1994}. Such PHs are typically difficult to find in the spectra of highly beamed sources such as blazars \citep{Aharonian2001}; therefore, we do not consider this component of observable $\gamma$-rays. The account of PHs would render the observable angular distribution even broader than in the framework of our model and thus would reinforce our conclusions. We also neglect synchrotron losses (see \citet{Aharonian2010} and Fig. 9 of \citet{Berezinsky2016} for related discussions).

The constraints on the strength of the EGMF in voids are being actively debated \citep{Neronov2010,Tavecchio2010,Dermer2011,Taylor2011,Arlen2014,Finke2015,Ackermann2018}; no reliable measurement of these magnetic fields is available by now. The typical energy loss length of cascade electrons is much smaller than the extension of the void (see \citet{Dzhatdoev2019b}, Fig. 2).\footnote{this figure was produced assuming the approximation for the IC process presented in \citet{Khangulyan2014}} Depending on the strength and the correlation length of the EGMF in voids, cascade $\gamma$-rays may form a PH or, in case of small electron deflection angle $\delta_{e} \ll 1$ rad, appear as a highly anisotropic ``magnetically broadened cascade'' (MBC) pattern \citep{Abramowski2014}. 

\subsection{Individual effect} \label{ssec:individual}

\begin{figure}
\centering
\includegraphics[width=8.4cm]{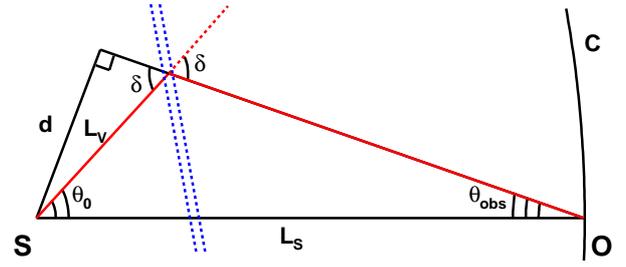}
\caption{Geometry scheme for the intergalactic hadronic cascade model and one filament (not to scale, see text for more details).}
\label{Fig01}
\end{figure}

Here we consider the case where the total deflection is due to one filament of the LSS\footnote{deflections on galaxy clusters are usually important for large redshifts, $z_{s} > 0.3-0.5$ (see Subsection 6.2 in D05)}. A simplified scheme of the corresponding geometry is shown in Fig.~\ref{Fig01}. A source ($S$) emits UHE protons that first propagate through a void (underdense region of space) with the diameter $L_{V}$ and then are upscattered on a filament (denoted by twin dashed blue lines; the deflection angle is, again, denoted as $\delta$). An example of a proton path is shown as a red line ($\theta_{0}$ is the primary proton emission angle). The observer ($O$), located at a distance $L_{S}$ from the source, detects secondary $\gamma$-rays from cascades initiated by the proton. Primary protons are deflected by the Galactic magnetic field and are not registered from the same direction as the $\gamma$-rays. 

We extrapolate the observable $\gamma$-ray path until we have a triangle with a right angle. The side $d$ of that triangle can be represented by the following equation:
\begin{equation}
d = L_{V} \sin(\delta) = L_{S} \sin(\theta_{obs}), \label{eq2}
\end{equation}
where $\theta_{obs}$ is the observable angle (the angle between the direction from the source to the observer and the direction of the incoming observable $\gamma$-ray)\footnote{for $\gamma$-rays produced \textit{after} the proton was deflected on the filament}. Therefore, $\theta_{obs}$ could be estimated as follows:
\begin{equation}
\sin(\theta_{obs}) = \sin(\delta) \frac{L_{V}}{L_{S}}. \label{eq3}
\end{equation}
We note that the same expression is widely known for purely electromagnetic cascades \citep{Neronov2009}; the physical interpretation of the quantities entering into this expression is, however, very different.

Following \citet{Harari2016}, we estimate a typical astrophysically plausible value of the deflection angle on a single filament as
\begin{equation}
\delta \approx 1^{\circ} \frac{B}{1 \: nG} \frac{40 \: EeV}{E_{p}/Z} \frac{\sqrt{L_{B}l_{c}}}{1 \: Mpc}, \label{eq4}
\end{equation}

where $B$ is the magnetic field strength, $Z$ is the charge number of the primary particle (proton in our case), $L_{B}$ is the thickness of the filament and $l_{c}$ is the correlation length for that magnetic field. For $B = 1$ nG, $L_{B} = 1$ Mpc, $l_{c} = 1$ Mpc, we get $\delta = 1^{\circ}$ for $E = 40$ EeV and $\delta = 1.33^{\circ}$ for $E = 30$ EeV. Our estimates of $\theta_{obs}$ for the blazar 1ES 1101-232 ($z_{s} = 0.186$), $\delta = 1^{\circ}$ and several values of $L_{V}$ are presented in Table~\ref{Tab01} (first line). The typical extension of an IACT PSF is about $0.1^{\circ}$, thus, a part of observable $\gamma$-ray flux in our case will not fit into the point-like image of the source.

\begin{table}
\caption{Estimates of the observable angle $\theta_{obs}$ [$^{\circ}$] for models with different number of filaments.}             
\label{Tab01}
\begin{tabular}{| c | c c c c |}
\hline      
$L_{V}$/$L_{V_{2}}$/$L$ & 50 Mpc & 100 Mpc & 200 Mpc & 500 Mpc \\
\hline                    
 one filament	        & 0.064  & 0.13    & 0.25    & 0.64    \\
 two filaments	        & 0.073  & 0.16    & 0.37    & 2.1     \\
 n filaments	        & 0.016  & 0.045   & 0.13    & 0.51    \\
\hline                    
\end{tabular}
\end{table}

Below, in Sect.~\ref{sec:results}, we present results for a specific case of the isotropic angular distribution of protons. This greatly simplifies calculations, as all observable $\gamma$-ray events are collected from the observer sphere~$C$. 

\subsection{Collective effect}

\begin{figure}
\centering
\includegraphics[width=8.4cm]{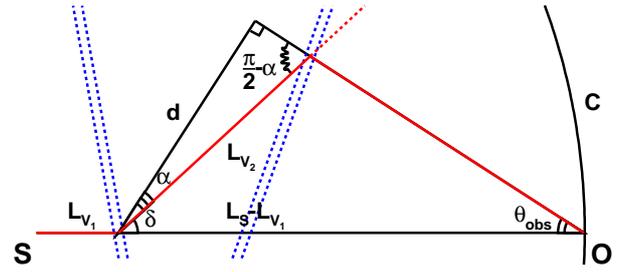}
\caption{Geometry scheme for the intergalactic hadronic cascade model and two filaments (not to scale).}
\label{Fig02}
\end{figure}

Now let us assume that the primary protons go through two filaments before reaching the observer (see Fig.~\ref{Fig02}). The notations here are mostly the same as in Subsect.~\ref{ssec:individual}, but we consider the specific case of $\theta_{0} = 0$ and introduce another void between the filaments with the diameter $L_{V_{2}}$ (the diameter of the first void is now denoted as $L_{V_{1}}$).

In analogy with eq.~(\ref{eq2})\footnote{here we consider $\gamma$-rays produced after the proton had undergone two deflections}:
\begin{equation}
d = (L_{S}-L_{V_{1}})\sin(\theta_{obs}) = L_{V_{2}}\sin\left(\frac{\pi}{2} - \alpha\right) \label{eq5}
\end{equation}
At the same time
\begin{equation}
\delta + \alpha = \frac{\pi}{2} - \theta_{obs} \rightarrow \frac{\pi}{2} - \alpha = \delta + \theta_{obs} \label{eq6}
\end{equation}
Assuming that $\theta_{obs} \ll \frac{\pi}{2}$ and $\delta \ll \frac{\pi}{2}$ we obtain the following:
\begin{equation}
(L_{S}-L_{V_{1}}) \theta_{obs} \approx L_{V_{2}}(\delta + \theta_{obs}), \label{eq7}
\end{equation}
and finally
\begin{equation}
\theta_{obs} \approx \frac{\delta L_{V_{2}}}{L_{S}-L_{V_{1}}-L_{V_{2}}}. \label{eq8}
\end{equation}
Assuming $L_{V_{1}} = 50$ Mpc and various values of $L_{V_{2}}$, we estimate $\theta_{obs}$ (see second line of Table~\ref{Tab01}). Finally, we consider the case of many filaments $n = L_{S}/L_{V} \gg 1$ when the total deflection of the primary proton $\delta \sim \delta_{n}\sqrt{n}$. For cascades initiated at the distance $L$ from the source:
\begin{equation}
\theta_{obs} \sim \delta_{n}\sqrt{\frac{L}{L_{V}}}\frac{L}{L_{S}} = \delta \left(\frac{L}{L_{S}}\right)^{3/2}. \label{eq9}
\end{equation}
Numerical estimates of $\theta_{obs}$ for this case are also presented in Table~\ref{Tab01} (third line).

\subsection{EM cascade spectral features}

\begin{figure}
\centering \vspace{1pc}
\includegraphics[width=8.4cm]{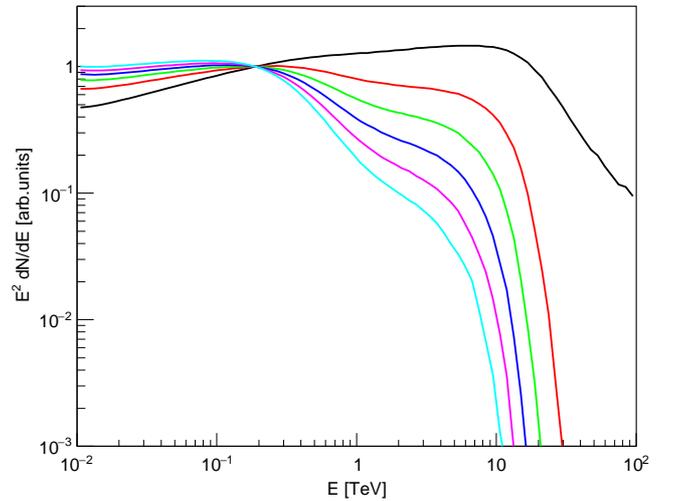}
\caption{SEDs of intergalactic EM cascades for different intervals of $z_{0}$. Black curve denotes the interval $0.0 < z_{0} < 0.03$, red --- $0.03 < z_{0} < 0.06$, green --- $0.06 < z_{0} < 0.09$, blue --- $0.09 < z_{0} < 0.12$, magenta --- $0.12 < z_{0} < 0.15$, cyan --- $0.15 < z_{0} < 0.18$. For convenience, all curves are normalized to unity at $E = 200$ GeV.}
\label{Fig03}
\end{figure}

The shape of the observable spectrum of an EM cascade depends on the distance between its origin and the observer. SEDs of cascades initiated by $\gamma$-rays with the energy $E_{0} = 1$ PeV are shown in Fig.~\ref{Fig03} for various ranges of their origin redshift $z_{0}$. These calculations were made with the ELMAG 2.03 publicly-available code \citep{Kachelriess2012}. The distribution on $z_{0}$ is random with a uniform probability density function. In Sect.~\ref{sec:results}--\ref{sec:prospects} of this work we present results for the two following EBL models: \citet{Kneiske2010} (hereafer KD10) and \citet{Gilmore2012} (hereafer G12); other options include \citet{Primack2005,Franceschini2008,Dominguez2011,Inoue2013,Stecker2016,Franceschini2017,Franceschini2018}. The KD10 option represents a lower limit for the EBL density; G12 is a realistic EBL model that is consistent with existing constraints \citep{Ackermann2012,Biteau2015,Abdalla2017,Korochkin2018,FermiLAT2018,Acciari2019c,Abeysekara2019}, at least for $z < 0.2$.

Qualitatively, the evolution of a well-developed intergalactic EM cascade (i.e. when the parent $\gamma$-ray or electron has a high enough energy $E_{0} > 100$ TeV for $z_{0} = 0.2$ or even higher for lower values of $z_{0}$, usually corresponding to several or more generations of cascade $\gamma$-rays and electrons) may be described as follows \citep{Berezinsky2011}: 1) the parent $\gamma$-ray is absorbed on a EBL or CMB photon, 2) then, the cascade develops very fast, 3) finally, cascade $\gamma$-rays reach the effective $\gamma\gamma$ absorption threshold on EBL $E(\tau=1)$. The observable spectrum carries an imprint of the pair production process, namely, the redshift-dependent high-energy cutoff, that is clearly visible in Fig.~\ref{Fig03} for all cases considered. The more distant the cascade origin is, the lower is the cutoff energy in the observable spectrum.

For the primary proton energy in excess of several EeV and below 100 EeV, parent electrons and $\gamma$-rays that initiate intergalactic cascades are mostly produced on CMB photons. In this case, for $E_{p} < 60/(1 + z)$ EeV the dominant energy loss process is the Bethe-Heitler pair production (BHPP), while for $60/(1 + z)$ EeV $< E_{p} < 100/(1 + z)$ EeV the photohadronic processes rapidly take over \citep{Berezinsky2006}. Furthermore, the typical energy range for photohadronic electrons and $\gamma$-rays is 1-50 EeV, while for BHPP electrons it is 100 TeV -- 1 EeV (see Figs. 2-3 in D17). For the range of parameters considered in the present work, the spectrum of observable $\gamma$-rays is practically independent of the energy and the type ($\gamma$-ray/electron) of the parent particle, but depends on $z_{0}$ (``weak universality'', see \citet{Berezinsky2016} and D17 (Appendix A)). For a discussion of the applicability of the weak universality assumption, the reader is referred to \citet{Dzhatdoev2017}; see also Subsect.~\ref{ssec:add-qual} and Subsect.~\ref{ssec:add-est} of the present paper.

\subsection{Discussion of estimates}

Above we have shown that the observable angular distribution broadens as the primary proton travels through the EGMF and, at the same time, the observable spectrum becomes harder for the case of cascades that were initiated relatively near to the observer (i.e. for the cascades characterised by comparatively low values of $z_{0}$). It is remarkable that the hard component of observable $\gamma$-rays with low $z_{0}$ at the same time has much broader angular distribution than the rest of observable $\gamma$-rays (see values in Table 1 and Fig.~\ref{Fig03}). This leads to a qualititive conclusion that the spectrum inside the PSF of the observing instrument appears to have a cutoff with respect to the spectrum integrated over all values of the observable angle $\theta_{obs}$. In the next four Sections we present a more precise calculation of this effect and discuss it in more details. 

\section{Simulations} \label{sec:simulations}

\subsection{Deflection of protons} \label{ssec:simulations-deflection}

We simulate the propagation of protons through the EGMF with the publicly available CRPropa3 code (Cosmic Ray Propagation Framework version 3) \citep{AlvesBatista2016}, accounting for all relevant particle interactions and calculating energy losses of these protons on every step, assuming the D05 EGMF model. The EGMF is represented by a $($132 $Mpc)^{3}$ volume simulated using the Quimby code \citep{Muller2016} and stored in an input file. Reflective boundary conditions were chosen; the trajectories were followed until they reach the observer's sphere. As a result of the proton propagation simulation, we obtained $10^{5}$ trajectories of protons with the primary energy $E_{p0} = 30$ EeV and starting points distributed randomly and uniformly inside a $($30 $Mpc)^{3}$ cube located in the central region of the volume. The length of each trajectory is $10^{3}$ Mpc \footnote{CRPropa3 allows to propagate particles to greater distances than the extension of the EGMF volume using the replication of the volume}. We conservatively consider the isotropic angular distribution of the protons.\footnote{for strongly anisotropic angular patterns of the primary proton beam, an additional defocusing effect is in place, which makes the observable spectrum even steeper}

\subsection{Observable $\gamma$-ray signal} \label{ssec:simulations-spectrum}

Following D17, we utilize a hybrid approach and combine semi-analytic calculations of proton energy losses during their propagation with statistical (Monte Carlo, MC) simulations of proton deflections (see the previous subsection), as well as MC calculations of intergalactic EM cascade spectra. Coordinates, directions and energies of the protons were calculated with the step $\delta z = 10^{-5}$. We accounted for pair production and pion production energy losses on CMB photons, as well as adiabatic losses (redshift) according to \citet{Berezinsky2006} (see Subsection 2.2 of D17). As an additional option we also included pion production energy losses on EBL photons; however, their impact is negligible for the range of parameters considered in the present work. We neglect intra-cluster and intra-filament light \citep{Ellien2019}.

The shape of the observable spectrum of $\gamma$-rays in the weak universality approximation is (D17, eq. 7):
\begin{equation}
\left(\frac{dN}{dE}\right)_{\gamma-obs}(E) \propto \int\limits_{0}^{z_{s}}{K_{em}(z)\frac{w(z)}{w(0)}\left(\frac{dN}{dE}\right)_{c}(E,z)dz} \label{eq10},
\end{equation}
where $K_{em}(z)$ is the fraction of ``active'' energy losses (i.e. pair production losses on the CMB, as well as pion production losses on the CMB and EBL) transferred to $\gamma$-rays and electrons, $(dN/dE_{c})(E,z)$ is the universal spectrum of EM cascade with a starting point at $z$, and $w(z)/w(0)$ is the energy transferred to $\gamma$-rays and electrons on the step $dz$, normalized to the value of $w$ at $z = 0$. Following \citet{Berezinsky2006}:
\begin{equation}
w(z) = \left|\left(\frac{dE_{p}}{dz}\right)_{pair+pion}\right| = \left|\left(\frac{dE_{p}}{dt}\right)_{pair+pion}\right|\left|\frac{dt}{dz}\right| \label{eq11}, 
\end{equation}
\begin{equation}
\left|\frac{dt}{dz}\right| = \frac{1}{H_{0}\sqrt{\Omega_{m}(1+z)^{3}+\Omega_{\Lambda}}}\frac{1}{1+z} \label{eq12}.
\end{equation}

Using the \mbox{ELMAG 2.03} code for the G12 EBL option and the 2.02 version of the same code for the KD10 EBL option \citep{Kachelriess2012}, we computed a database of the universal EM cascade spectra $(dN/dE_{c})(E,z)$ with the primary energy 1 PeV and $z$ distributed randomly and uniformly varying from 0 to 0.30 and use this database to compute the observable spectrum according to eq.~(\ref{eq10}). Other up-to-date publicly-available intergalactic cascade codes have more advanced functionality than \mbox{ELMAG~2.03}: namely, \citet{Kalashev2015} enables the account of the universal radio background (URB), and \citet{Fitoussi2017}, as well as the new version of ELMAG (3.01) \citep{Blytt2019} allow to perform realistic simulation of three-dimensional (3D) structure of intergalactic EM cascades. However, in the scope of the present paper the \mbox{ELMAG 2.03} code is suitable for our purposes (which are mainly to calculate the energy spectra of intergalactic EM cascades), at the same time retaining significant advantages of greater simplicity, reliability, and speed compared to the more advanced codes.

In our simulation framework the observer is represented by a sphere (denoted as $C$ in Fig.~\ref{Fig01} and Fig.~\ref{Fig02}) with the source in its center, the comoving distance from the source to the observer being the radius of the sphere. Observable $\gamma$-rays are collected at the intersection with the sphere. Thus, the spherical symmetry of our setup allows to greatly simplify the calculations for the case of the isotropic angular distribution of protons. 

Finally, we compute and tabulate the observable angle $\theta_{obs}$ for every proton trajectory simulated as was described in Subsect.~\ref{ssec:simulations-deflection}. These tables have the step of 5 Mpc on the traversed distance; in what follows we use them to calculate the observable $\gamma$-ray angular distribution.

\section{Results} \label{sec:results}

\subsection{Observable spectrum with and without the account of proton deflection} \label{ssec:results-spectrum}

In order to illustrate the original idea of \citet{Uryson1998}\footnote{many works, including \citet{Essey2010,Essey2011,Zheng2015,Archer2018} are basically founded on the same idea as the one proposed by \citet{Uryson1998}}, in this Section we present results with and without the account of primary proton deflection and compare them.

In D17 we called the version of the IHCM neglecting proton deflection ``the basic intergalactic hadronic cascade model'' (BHCM) and calculated the observable spectra for a set of sources using this model. Now we repeat the calculation of the observable spectrum, taking the observable angle into consideration. We retain only those observable $\gamma$-rays that have $\theta_{obs} < \theta_{PSF} = 0.1^{\circ}$ and reject those with $\theta_{obs}\ge\theta_{PSF}$. Results for a more realistic PSF are presented below in Subsect~\ref{ssec:results-psf}. In D17 we had already shown that the effect of proton deflection strongly modifies the observable spectrum, therefore, we called this version of the IHCM ``the modified intergalactic hadronic cascade model'' (MHCM). 

\begin{figure}
\centering \vspace{1pc}
\includegraphics[width=8.4cm]{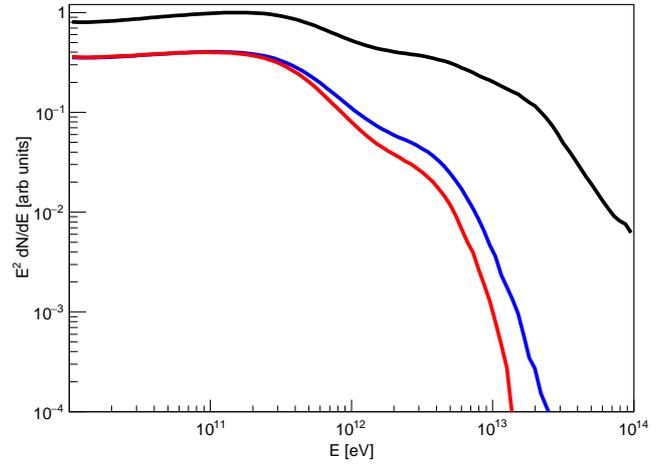}
\caption{Observable SEDs for intergalactic cascade models for $z_{s} = 0.186$ and the KD10 EBL option. Black curve denotes an SED for the basic hadronic cascade model, blue curve --- an SED for the modified hadronic cascade model. Red curve denotes an SED for the intergalactic electromagnetic cascade model (``the universal SED'').}
\label{Fig04}
\end{figure}

Here we present, for the first time, the results of detailed calculations based on a model of LSS formation predicting a highly inhomogeneous EGMF, namely, the D05 model. They are presented in Fig.~\ref{Fig04} in the form of an observable SED (blue curve). The SEDs for the BHCM (black curve) and the elecromagnetic cascade model (ECM) in the universal regime (red curve) are presented as well; all three curves were calculated assuming the KD10 EBL option\footnote{the SED of the BHCM presented in Fig.~\ref{Fig04} was normalized to unity at its maximal value; the SED for the ECM in the universal regime was normalized to the SED of the MHCM at $E = 100$~MeV}.

As one can see from Fig.~\ref{Fig04}, the observable spectrum in the framework of the BHCM is much harder than the universal EM cascade spectrum. As was explained in the Introduction, the reason behind this effect is that a part of EM cascades initiated by the protons have the redshift of the origin $z_{0} \ll z_{s}$, and so these cascades have much harder spectra than those with $z_{0} = z_{s}$ (see also Fig.~\ref{Fig03} and the related discussion). To put it simply, in the framework of the BHCM the source becomes effectively much nearer to the observer, reducing the impact of intergalactic $\gamma\gamma$ absorption on the observable spectrum.

After the account of proton deflection, however, the situation appears to be very different with respect to the oversimplified case of the BHCM. We note that there is a marked difference between the shapes of the corresponding SEDs in the framework of the basic and modified hadronic models. The MHCM SEDs have a high-energy cutoff at $E \sim 10$ TeV. This cutoff is due to the relatively broad angular distribution of observable $\gamma$-rays from cascades initiated near the observer. The results presented in Fig.~\ref{Fig04} very well match our expectations based on the estimates made in Sect.~\ref{sec:estimates}. 

\begin{figure}
\centering \vspace{1pc}
\includegraphics[width=8.4cm]{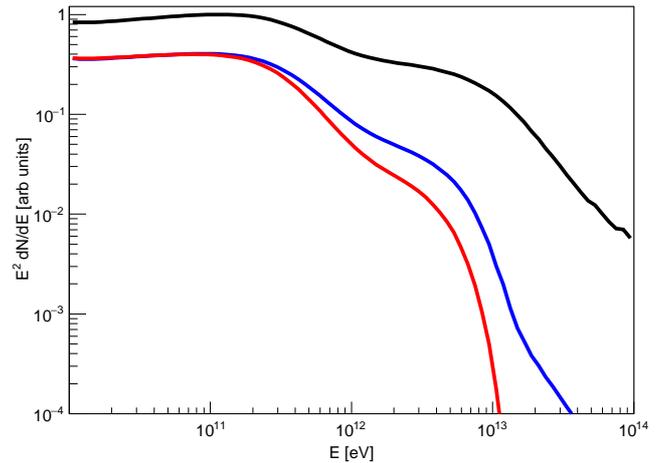}
\caption{The same as in Fig.~\ref{Fig04}, but for the G12 EBL option.}
\label{Fig05}
\end{figure}

In Fig.~\ref{Fig05} we present the results of a calculation similar to the one presented in Fig.~\ref{Fig04}, but done assuming the G12 EBL option. Observable SEDs for the case of the MHCM and the KD10 EBL model option were calculated by averaging over $10^{4}$ proton trajectories and $10^{3}$ proton trajectories were averaged for the G12 EBL option. We have found that averaging over $10^{3}$ trajectories is sufficient to obtain stable enough results for $E < 30$ TeV (see Appendix~\ref{appendixa}, Fig.~\ref{FigA1} for details). Observable $\gamma$-ray spectra for $10^{2}$ individual trajectories are also presented in Appendix~\ref{appendixa} (see Fig.~\ref{FigA2}). We also note that the difference between the spectra of the modified hadronic models and the universal spectra of purely EM cascades is comparable to the difference between these universal spectra for various EBL options.

It is useful to introduce a new parameter called ``the critical distance'' $L_{cr}$ defined as the distance from the observer at which the observable angle is equal to the typical extension of the PSF (e.g. the 68 \% containment angle): $\theta_{obs} = \theta_{PSF}$. A histogram of the distribution on $L_{cr}$ for $\theta_{PSF} = 0.1^{\circ}$ is shown in Appendix~\ref{appendixb} (see Fig.~\ref{FigB1})\footnote{here $L_{cr}$ was calculated for individual trajectories}. This histogram is peaked at $L_{cr} \approx 200$~Mpc with only a small minority of trajectories having $L_{cr} > 300$~Mpc.

The relative contributions of extragalactic filaments and clusters to the total proton deflection vary with the redshift of the source. A border between the clusters and the filaments on the plane of the variables ``deflection angle -- the cumulative fraction of the sky with deflection angle larger than some threshold'' is shown in Fig. 15 of D05 as dot-dashed magenta line. We estimate that for $z = 0.186$ the fraction of the sky covered by clusters is $\approx 0.2$. For $z = 0.3-0.4$ this fraction rises to 0.5; therefore, for larger $z$ the proton deflection is typically dominated by galaxy clusters. Additionally, the typical deflection in a cluster is significantly larger than the one in a filament \citep{Murase2012}. We note that the results of such estimates are strongly dependent on a particular line-of-sight chosen.

\begin{figure}
\centering \vspace{1pc}
\includegraphics[width=8.4cm]{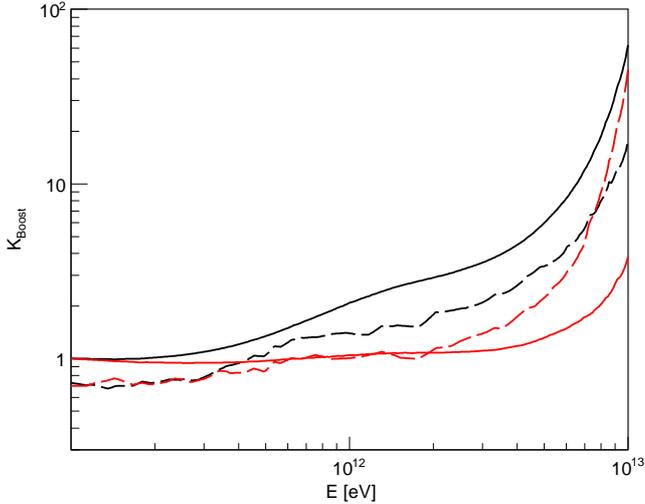}
\caption{Flux boost factors for the BHCM (black solid curve), MHCM (red solid curve), and two models including the $\gamma \rightarrow ALP$ oscillation effect with $\xi = 5.0$ (black dashed curve) and $\xi = 0.5$ (black dashed curve).}
\label{Fig-Boost}
\end{figure}

Finally, we calculated the so-called flux boost factors $K_{Boost}(E)$ \citep{SanchezConde2009}, defined as the ratio between the spectrum in the framework of a certain model to the spectrum in the framework of the absorption-only model. For BHCM and MHCM, the ``absorption-only'' spectrum was assumed according to Fig. 12 of \citet{Dzhatdoev2017} (top-left) and the G12 EBL model (denoted as black curve in this figure). Additionally, we calculated $K_{Boost}(E)$ for two models including the $\gamma \rightarrow ALP$ oscillation effect according to \citet{DeAngelis2011}, Fig. 3 (second row, left) that were developed by the authors of this paper for $z = 0.188$, the EGMF coherence length of 4 Mpc, and two values of $\xi \equiv (B_{0}\ [nG])(10^{11}\ [GeV]/M)$ (here $M = 1/g$ and $g$ is the ALP-photon coupling constant; \citet{DeAngelis2011} assumed the EGMF strength evolution $B = B_{0}(1+z)^{2}$).\footnote{\citet{DeAngelis2011} assumed the EBL model of \citet{Franceschini2008} that for $z < 0.2$ predicts the EBL parameters close to those predicted by the G12 model} Flux boost factors in the energy range from 100 GeV to 10 TeV for the four abovementioned models are shown in Fig.~\ref{Fig-Boost}.

For the basic hadronic model the modification factor is comparable with those for both of $\gamma \rightarrow ALP$ models. Therefore, in the framework of the BHCM, intergalactic cascades initiated by the primary protons indeed represent a dangerous source of background in ALP searches. For the MHCM, however, the modification factor at $E > 3$ TeV is significantly smaller than for both of $\gamma \rightarrow ALP$ models, allowing to distinguish between these models.

\subsection{Observable spectrum: the case of a realistic PSF} \label{ssec:results-psf}

\begin{figure}
\centering \vspace{1pc}
\includegraphics[width=8.4cm]{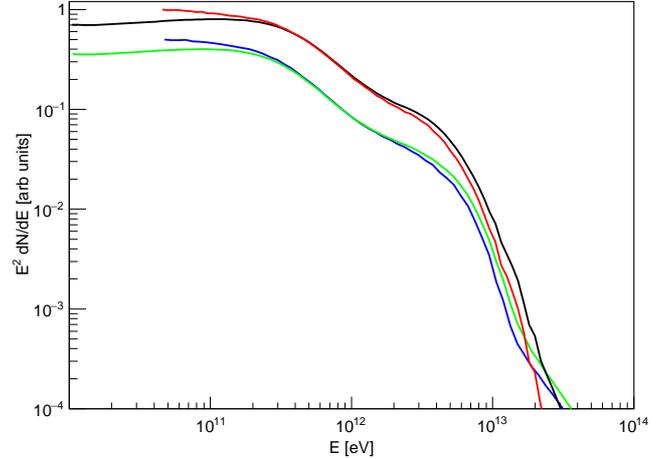}
\caption{A comparison between the SEDs for a simplified PSF with $\theta_{PSF} = 0.1^{\circ}$ and a realistic one (for the H.E.S.S. array) in the MHCM. Black curve denotes an SED for a simplified PSF and the KD10 EBL option, green curve --- the same for the G12 EBL option. Red curve denotes an SED for the H.E.S.S. PSF and the KD10 EBL option, blue curve --- the same for the G12 EBL option. SEDs with different EBL options are spaced apart by the factor of two for illustrative purposes.}
\label{Fig06}
\end{figure}

The width of the PSF for most of the contemporary $\gamma$-ray detectors depends on the energy. This impacts the shape of the observable SED in the framework of the MHCM. In Fig.~\ref{Fig06} we compare SEDs for a simplified representation of a PSF with $\theta_{PSF} = 0.1^{\circ}$ and a more realistic PSF for the case of the H.E.S.S. IACT array \citep{CTA2018}. The SEDs for both cases were averaged over $10^{4}$ trajectories for the KD10 EBL option and over $10^{3}$ trajectories for the G12 option.

The H.E.S.S. PSF becomes wider at lower energies, so the intensity at low energies appears to be greater than the one for the case of $\theta_{PSF} = 0.1^{\circ}$. Thus, for the realistic PSF the observable spectrum becomes steeper than for the case of an energy-independent PSF.

\subsection{Observable spectrum: comparison with observations} \label{ssec:results-comp}

\begin{figure}
\centering \vspace{1pc}
\includegraphics[width=8.4cm]{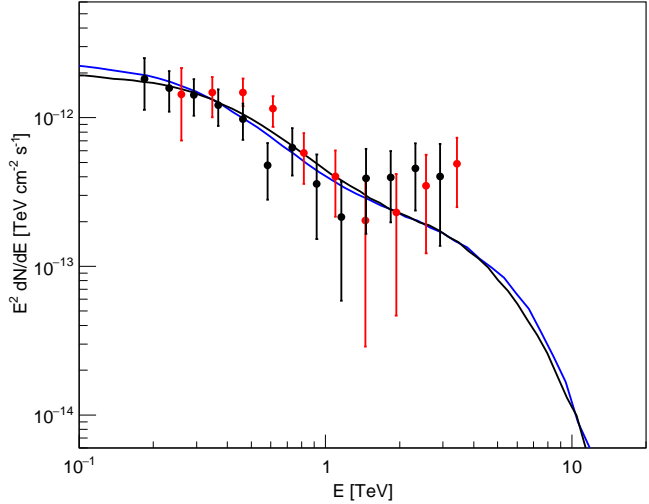}
\caption{SEDs of the source 1ES1101-232 measured with the H.E.S.S. IACT array: circles denote measurements, bars --- their uncertainties; the results of the 2006 analysis are shown in black, the results of the 2007 reanalysis --- in red. Black curve denotes the SED for the MHCM assuming the H.E.S.S. PSF and the KD10 EBL option, blue curve --- the same, but for the G12 EBL option.}
\label{Fig07}
\end{figure}

In Fig.~\ref{Fig07} we show an SED of the extreme TeV blazar \mbox{1ES 1101-232} ($z_{s} = 0.186$) measured by the H.E.S.S. Collaboration \citep{Aharonian2006}. In 2007 the H.E.S.S. Collaboration had performed a reanalysis of the same dataset \citep{Aharonian2007b}; the resulting SED is also presented in Fig.~\ref{Fig07} for comparison. In the same figure we show two model curves in the framework of the MHCM for various EBL models; the realistic PSF of the H.E.S.S. IACT array was assumed for both model options.\footnote{these model curves were shown in Fig.~\ref{Fig06} (see Subsect.~\ref{ssec:results-psf} for more details); here we renormalized them to fit the measured SEDs}

\begin{figure}
\centering \vspace{1pc}
\includegraphics[width=8.4cm]{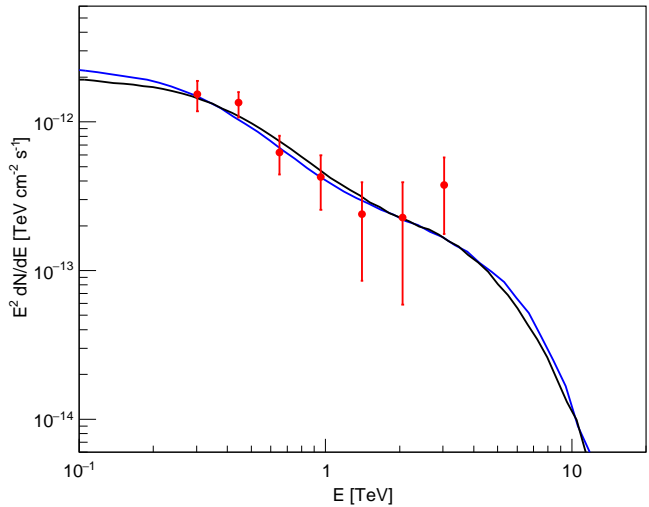}
\caption{The same as in Fig.~\ref{Fig07}, but the measurements are shown for the source 1ES 0347-121.}
\label{Fig08}
\end{figure}

In Fig.~\ref{Fig08} we show an SED of the extreme TeV blazar \mbox{1ES 0347-121} ($z_{s} = 0.188$) that was also measured by the H.E.S.S. Collaboration \citep{Aharonian2007c}, and present the same two model curves as in Fig.~\ref{Fig07}.\footnote{the small difference of the measured redshift between \mbox{1ES 1101-232} and \mbox{1ES 0347-121} allows one to use the same models} For both blazars, the figures demonstrate a reasonable agreement between the observed SEDs and the model ones, except for a slight excess in the measured SEDs at the highest energies.

\subsection{Observable angular distribution} \label{ssec:results-ang}

Besides the energy spectrum, $\gamma$-ray detectors can also measure the angular distribution of the source. In particular, the {\it \mbox{Fermi-LAT}} Collaboration have recently performed a search for spatial extension in some high-latitude ($|b| > 5^{\circ}$) sources, including the blazars \mbox{1ES 1101-232} and \mbox{1ES 0347-121} \citep{Biteau2018}. They have drawn the conclusion that the images of these sources are compatible with the ones expected from point-like sources. The H.E.S.S. Collaboration have presented the same conclusion in \citet{Aharonian2007c} for \mbox{1ES 0347-121}; the observable angular distribution for \mbox{1ES 1101-232} also does not show any obvious hints at an additional extension \citep{Aharonian2007b}.

\begin{figure}
\centering \vspace{1pc}
\includegraphics[width=8.4cm]{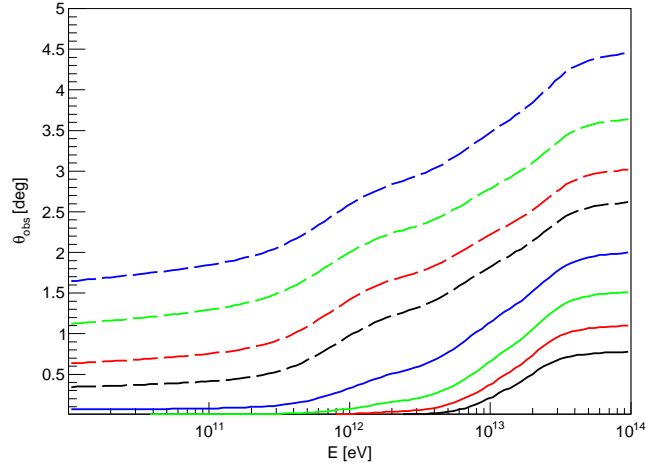}
\caption{Quantiles of the observable angular distribution for the KD10 EBL model option: black solid curve denotes a 5$\%$ containment angle, red solid curve --- 10$\%$ containment angle, green solid curve --- 20$\%$, blue solid curve --- 40$\%$, black dashed curve --- 68$\%$, red dashed curve --- 80$\%$, green dashed curve --- 90$\%$, and blue dashed curve --- 95$\%$.}
\label{Fig09}
\end{figure}

In Fig.~\ref{Fig09} we show a model of the observable angular distribution that could be compared with observations; this figure demonstrates how various containment values of the observable angle $\theta_{obs}$ change with the energy.\footnote{here we work with the angular distribution truncated at 10$^{\circ}$; this value represents the diameter of the field-of-view of a typical IACT or a typical extension of the region-of-interest (ROI) in {\it \mbox{Fermi-LAT}} data analysis} We note that there is a steady increase in the value of $\theta_{obs}$ as the observable energy increases, with $\approx 32$\% of observable angles being larger than 1$^{\circ}$ at $E > 1$~TeV, and $\approx 60$ \% of observable angles exceeding this value at $E > 8$~TeV.

Thus, the intergalactic hadronic cascade model for the blazars \mbox{1ES 1101-232} and \mbox{1ES 0347-121} predicts observable spectra similar to those measured by the H.E.S.S. Collaboration, but this model is not favoured in view of observed angular distributions typical for point-like sources. A detailed statistical analysis supporting this conclusion is in preparation; it will be published elsewhere. Here we conclude that measurement of the observable angular distribution would provide a critical test for the validity of the intergalactic hadronic cascade model.

\section{Observational prospects} \label{sec:prospects}

\begin{figure}
\centering \vspace{1pc}
\includegraphics[width=8.4cm]{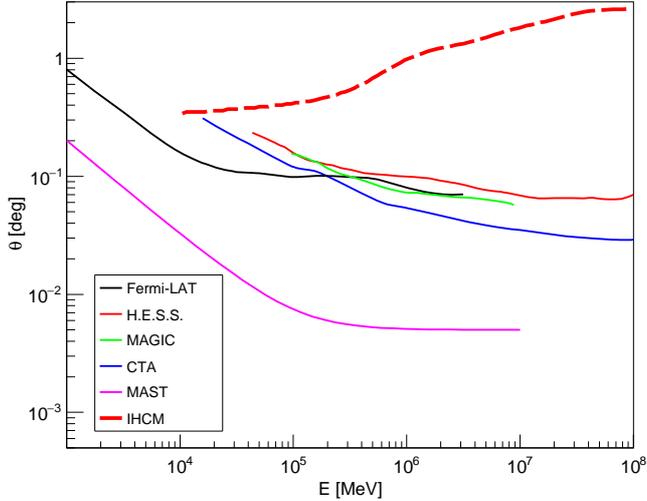}
\caption{Angular resolution (68 \% containment angle) of various gamma-ray instruments (solid curves: black curve denotes {\it \mbox{Fermi-LAT}}, red curve --- \textit{\mbox{H.E.S.S.}}, green curve --- \textit{\mbox{MAGIC}}, blue curve --- \textit{\mbox{CTA}}, magenta curve --- {\it \mbox{MAST}}) vs. the 68 \% containment angle of the observable emission assuming the MHCM with the KD10 EBL.}
\label{Fig10}
\end{figure}

In this section we compare the 68$\%$ containment radius of the simulated angular distribution with the same quantity for the PSF of several operating and projected $\gamma$-ray telescopes (see Fig.~\ref{Fig10}). In this figure we present the single-photon angular resolution $\theta_{68}(E)$ for {\it \mbox{Fermi-LAT}} according to \citet{Atwood2009};  $\theta_{68}(E)$ for \textit{\mbox{H.E.S.S.}}, \textit{\mbox{MAGIC}} and \textit{\mbox{CTA}} are taken from \citep{CTA2018}. The angular resolution for \textit{\mbox{VERITAS}} \citep{Krennrich2004,Park2016} is similar to that of \textit{\mbox{H.E.S.S.}} and \textit{\mbox{MAGIC}}. In addition, we show the angular resolution for the {\it \mbox{MAST}} space $\gamma$-ray telescope proposed in \citet{Dzhatdoev2019a}. The {\it \mbox{MAST}} concept (an abbreviation for ``Massive Argon Space Telescope'') is based on the liquid Argon time projection chamber (TPC) approach \citep{Dolgoshein1970,Rubbia1977,Rubbia2011}. Such a detector with the total sensitive mass of $\approx 36$~t could be launched with the latest contemporary launch vehicles such as Falcon Heavy.

Finally, in Fig.~\ref{Fig10} we also present the 68 \% containment angle of the observable emission in the framework of the MHCM. We note that this extended emission is readily discernible by all considered $\gamma$-ray telescopes at $E > 10$ GeV provided that the exposure is sufficiently high and the background is sufficiently low. Next-generation IACTs such as \textit{\mbox{CTA}} are especially well-suited for the search of broadened patterns around the positions of presumably point-like sources as the extension of the model angular distribution significantly increases with the energy. Projected space $\gamma$-ray telescopes with improved angular resolution, broad field-of-view, and high duty cycle, such as {\it \mbox{MAST}}, could also reveal the extended nature of these sources.

\begin{figure}
\centering \vspace{1pc}
\includegraphics[width=8.4cm]{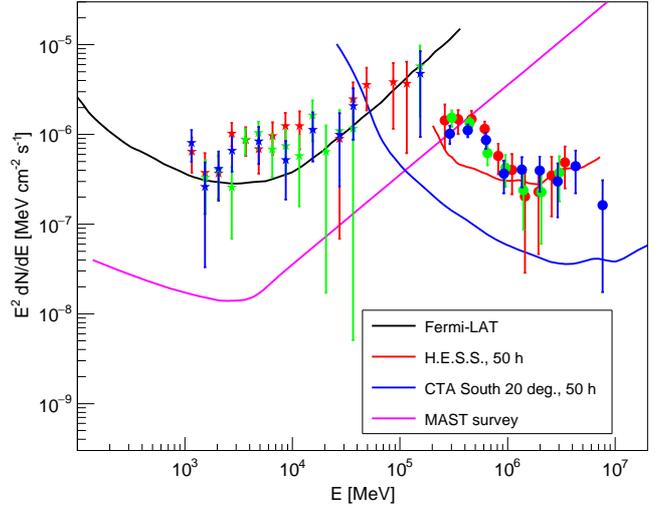}
\caption{SEDs of some extreme TeV blazars measured with IACTs (circles) and {\it \mbox{Fermi-LAT}} (stars) together with sensitivities of several $\gamma$-ray telescopes. Red symbols denote \mbox{1ES 1101-232}, green symbols --- \mbox{1ES 0347-121}, blue symbols --- \mbox{1ES 0229+200}; black curve denotes the sensitivity of {\it \mbox{Fermi-LAT}}, red curve --- \textit{\mbox{H.E.S.S.}}, blue curve --- \textit{\mbox{CTA}}, magenta curve --- {\it \mbox{MAST}}.}
\label{Fig11}
\end{figure}

In Fig.~\ref{Fig11} we show the SEDs of extreme TeV blazars \mbox{1ES 1101-232}, \mbox{1ES 0347-121}, and \mbox{1ES 0229+200} ($z_{s} = 0.14$) measured with \textit{\mbox{H.E.S.S.}} \citep{Aharonian2007a,Aharonian2007c}, \textit{\mbox{VERITAS}} \citep{Aliu2014}, and {\it \mbox{Fermi-LAT}} \citep{Biteau2018} in comparison with the differential sensitivity for point-like sources of various $\gamma$-ray telescopes. The sensitivity for \textit{\mbox{MAGIC}} and \textit{\mbox{VERITAS}} is qualitatively similar to that of \textit{\mbox{H.E.S.S.}} The sensitivity of {\it \mbox{Fermi-LAT}} is barely sufficient to measure the spectra of these sources; however, IACTs can readily detect extreme TeV blazars due to their very hard intrinsic spectra. Next-generation space $\gamma$-ray telescopes such as {\it \mbox{MAST}} could also measure the spectra of such sources due to improved sensitivity.

\section{Discussion} \label{sec:discussion}

For the calculations performed in the present paper we assumed specific model parameters, including the source redshift $z_{s} = 0.186$, the primary proton energy $E_{p0} = 30$ EeV, the D05 model for the EGMF, and two models of the EBL (namely, KD10 and G12). Here we discuss how the results of our calculations would change if we assume other values of these parameters or account for ``additional processes'', such as triplet pair production (TPP, $e\gamma \rightarrow 3e$), double pair production (DPP, $\gamma \gamma \rightarrow 4e$), and interactions on the URB.

\subsection{Source redshift \label{ssec:redshift}}

At redshifts higher than $z_{s} = 0.3$ the fraction of proton trajectories that experience deflections on the EGMF of galaxy clusters becomes non-negligible, significantly increasing the typical deflection angle and making the effect of the angular broadening and the associated cutoff in the observable spectrum even more pronounced. Conversely, at relatively low redshifts ($z_{s} < 0.1$) a significant deviation of the observable spectrum in the framework of the MHCM from that of the universal spectrum of a purely EM cascade is expected. On the other hand, the width of the observable angular distribution rises with increasing energy, while the angular resolution of IACTs becomes better with the energy (see Fig.~\ref{Fig09} and Fig.~\ref{Fig10}), making the angular extensions of the sources readily discernible, even for the case of low redshifts.

\subsection{Primary proton energy \label{ssec:energy}}

For the case of the primary proton energy lower than 30 EeV, the primary protons deflect even stronger and the width of the observable angular distribution increases still further. The same is true for primary nuclei with the primary energy $< 30\cdot Z$ EeV. We note that the results obtained with the Pierre Auger Observatory disfavour the hypothesis of a purely proton composition above the energy of 10 EeV \citep{Abraham2010,Aab2014,Aab2017}.


For the case of high $E_{p0} > 100$ EeV, the characteristic energy loss length $L_{p-E}$ is lower than 100 Mpc \citep{Berezinsky2006}. Therefore, at first, protons lose energy rapidly until they reach the effective threshold of the pion photoproduction process ($\sim 60/(1+z)$ EeV). Because of this, the width of the angular distribution at high energies ($E > 1$~TeV) becomes only slightly smaller for $E_{p0} \gg 30$ EeV as compared to the case of $E_{p0} = 30$~EeV if the source is sufficiently distant, $z_{s} > 0.1$.

\subsection{EGMF model \label{ssec:egmf-model}}

The D05 EGMF model is usually considered to be conservative in terms of the predicted proton deflection, i.e. it predicts proton deflection to be smaller than the predictions of other models such as \citet{Sigl2004,Hackstein2018} (see \citet{Eichmann2019}, Subsect. 2.2). An attempt to reduce the angular extension of the observable emission beyond the observability level for contemporary experiments such as the \textit{\mbox{CTA}} array would require a significant revision of the existing models of the EGMF in filaments and clusters of the large scale structure.

First detections of synchrotron emission from extragalactic filaments have just started to appear \citep{Govoni2019,Vacca2018}. Based on the {\it \mbox{LOFAR}} observations reported in \citet{Govoni2019}, \citet{Brunetti2020} derived $B = 500-600$ nG for the filament connecting the galaxy clusters Abell 0399 and Abell 0401, much in excess of the typical $B$ in the framework of the D05 model. It is likely that the filaments detected in \citet{Govoni2019,Vacca2018} represent only a ``tip of the iceberg'' of the more numerous population of extragalactic filaments with a weaker magnetic field but a higher total volume filling factor. Future observations with more sensitive instruments, such as the {\it \mbox{SKA}} array \citep{Heald2020}, would likely reveal these fainter filaments. Concerning the assumptions and uncertainties of EGMF simulations, the reader is referred to \citet{Sigl2004,Dolag2005,Vazza2017,Hackstein2018,AlvesBatista2019}. Results assuming a more up-to-date EGMF model compared to D05 are in preparation and will be published elsewhere.

The ratio of the synchrotron and Compton luminosities of electrons in a filament is $P_{S}/P_{IC} \sim (B/B_{cmb})^{2}$ \citep{Brunetti2020}, where $B_{cmb} = 3.25$ $\mu$G$\cdot (1+z)^{2}$.\footnote{indeed, the number of Compton scattering acts is $\propto(1+z)^{3}$, and the average energy of produced $\gamma$-rays is $\propto(1+z)$} Therefore, detecting intergalactic filaments by observing their synchrotron radiation is difficult because for $B < 1$ $\mu$G most of the radiated power is transferred to Compton photons rather than to synchrotron ones. This could explain the small number of intergalactic filaments with synchrotron radiation detected so far, which results in the current low number of EGMF measurements in these filaments. Observations with X-ray missions such as {\it \mbox{XMM-Newton}} \citep{Jansen2001}, {\it \mbox{Chandra}} \citep{Chandra2019}, {\it \mbox{SWIFT}} \citep{Gehrels2004}, {\it \mbox{SRG}} \citep{Merloni2012} as well as with the proposed {\it \mbox{Athena}} X-ray observatory \citep{Nandra2013} could provide additional constraints on the value of $B$ or even result in the measurement of the EGMF strength in some extragalactic filaments.

\subsection{EBL model \label{ssec:ebl-model}}

The intensity of the EBL relevant for primary $\gamma$-ray absorption in the observable energy range $E < 20$ TeV is believed to be well-constrained now \citep{Ackermann2012,Abdalla2017,FermiLAT2018,Acciari2019c,Abeysekara2019,Franceschini2019,Biasuzzi2019}, with characteristic uncertainty of the order of tens of percent at $z < 0.5$. In the present paper we have shown the results of calculations assuming the KD10 EBL model that represents a lower limit for the EBL intensity. Other EBL models with higher intensities would result in an additional steepening of the observable spectrum with respect to the KD10 model one.

\subsection{Angular distribution of primary protons \label{ssec:angular}}

For simplicity, in the present paper we assumed the isotropic angular distribution of primary protons. Therefore, the results presented here are relevant for the case of $\theta_{p} \gg \delta \sim 1^{\circ}$, where $\theta_{p}$ is the width of the angular distribution of primary protons. For $\theta_{p}\le\delta$, instead, an additional effect of the beam defocusing is in place, for which the intensity at high energies is diminished further, thus, making the high-energy cutoff in the observable spectrum even more pronounced.

\subsection{The impact of URB, TPP, and DPP on the observable spectrum \label{ssec:add-proc}}

\subsubsection{Qualitative discussion \label{ssec:add-qual}}

\begin{figure}
\centering \vspace{1pc}
\includegraphics[width=8.4cm]{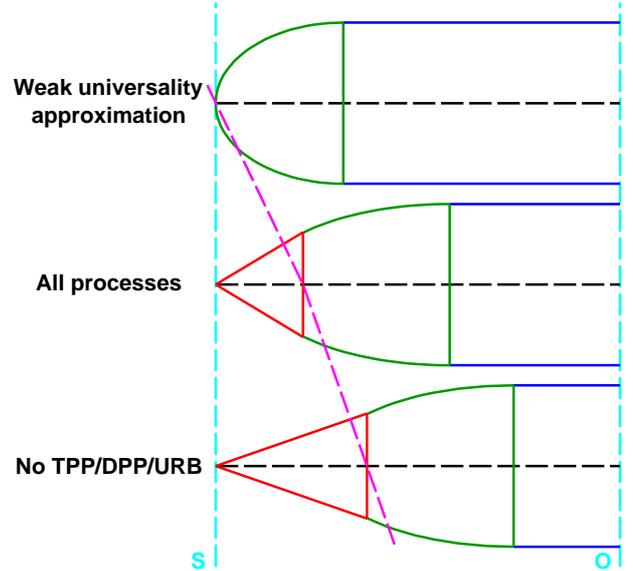}
\caption{Simplified scheme of geometry of electromagnetic cascade. Three main development stages of the cascade are shown: the high energy stage, the particle multiplication stage, and the low energy stage. Red cones denote the high energy stage, green parabolic segments --- the particle multiplication stage, blue parallel lines --- the low energy stage. Top panel denotes the weak universality approximation, middle panel --- the case when all processes are taken into consideration, bottom panel --- the case when the additional processes are neglected. S denotes the source, O --- the observer. To the right of the dashed  magenta line the additional processes could be safely neglected. We note that the length of the high energy stage is lower when all processes are accounted for compared to the case when the additional processes are neglected (see the text for more details).}
\label{Fig12}
\end{figure}

Interactions on the URB as well as the TPP and DPP processes can have an impact on the observable spectrum. URB intensity estimates were obtained in \citet{Protheroe1996,Niu2021}. The TPP process was considered in \citet{Jarp1973,Mastichiadis1986}, the DPP process --- in \citet{Brown1973,Lee1998}. \citet{Mastichiadis1994} investigated the consequences of including TPP into calculations of EM cascade development. They reached the general conclusion that ``the importance of triplet production decreases as the path length increases'' and that TPP ``has no effect on saturated cascades'', i.e. cascades with $\gamma$-rays below the pair production threshold.

Since the ELMAG code, which we use in this work to calculate cascades, does not account for these additional processes, below we aim to put the results of \citet{Mastichiadis1994} firmly within the context of the present work and justify the use of ELMAG.

Let us consider various stages of EM cascade development (see Fig.~\ref{Fig12}). The top panel of this figure corresponds to the case of the weak universality approximation. In this case the cascade enters the ``particle multiplication stage'' immediately, i.e. the energy of particles is divided roughly in equal proportions between the secondaries. This corresponds to the case of the primary $\gamma$-ray/electron energy $E_{0} = 1-10$ PeV. In two other cases, the cascade may first enter the ``high energy stage'' if the primary energy is high enough ($E_{0} > 100$~PeV); in this stage one of the secondaries may carry almost all of the primary energy. Finally, the last stage of cascade development (after the particle multiplication stage) is the ``low energy stage''; at this stage the remnant cascade photons gradually become ``sterile'' \citep{Berezinsky2016}, i.e. their pair production mean free path becomes much greater than the distance from the source to the observer. These stages were discussed in \citet{Berezinsky2016}.

Dashed magenta line in Fig.~\ref{Fig12} denotes the start of the particle multiplication stage. After this stage is set, the additional processes could be safely neglected, because the energy of the cascade particles at this point is typically below 10 PeV. We note that the spectral universality is the property of the particle multiplication stage. If the primary energy is high enough, then the cascade first enters the high energy stage, and the universality assumption is, to some extent, violated, even if no additional processes, such as the interactions on the URB as well as the TPP and DPP processes, are accounted for.

The extent to which the universality assumption is violated is defined by the following parameter: the ratio of the typical length of the high energy stage to the distance from the source to the observer. In the next subsection we show that the typical length of the high energy stage is actually lower for the case when the additional processes are accounted for compared to the case when these additional processes are neglected. Therefore, the observable spectra obtained in the ``weak universality'' approximation are actually less different from the spectra calculated with the account of all relevant processes than the universal spectra compared to the spectra without the account of the additional processes.

\subsubsection{Estimates \label{ssec:add-est}}

Let us consider the development of intergalactic EM cascades for the case of the primary particle energy $E_{0} > 100$ PeV, at first neglecting the interactions on the URB as well as the TPP and DPP processes. In this case the average fraction of the primary energy lost in each collision (usually called ``average inelasticity'') is (e.g. eq. (4) of \citet{Berezinsky2016}):
\begin{equation}
f_{in} \approx \frac{1}{ln \left(2x \right)} \label{eq13},
\end{equation}
where the parameter $x = E_{0}E_{b}/(m_{e}c^{2})^{2}$ is typically $\gg 1$ (here $m_{e}$ is the mass of electron and $E_{b}$ is the background photon energy). Therefore, one of the secondaries typically carries almost all of the primary energy, and the cascade is in the high-energy or leading-particle stage: $\gamma \rightarrow e \rightarrow \gamma' \rightarrow e'$ \citep{Berezinsky2016}. 

In this paper we mainly consider the case of the primary proton energy $E_{p0} \le 30$ EeV (see Subsect.~\ref{ssec:energy} above). For moderate redshifts $z < 0.5$ such protons lose energy mainly through pair production; the typical energy of the produced electrons is below 1 EeV (see e.g. Fig. 2 of \citet{Dzhatdoev2017}). The mean free path for $\gamma$-rays and electrons with the energy between 100 PeV and 100 EeV could be estimated as:
\begin{equation}
\lambda_{\gamma,e} \approx C_{\gamma,e} \left( \frac{E_{0}}{100 \: PeV} \right)^{0.85} \label{eq14},
\end{equation}
where for $\gamma$-rays $C_{\gamma} = 50$ kpc, and for electrons $C_{e} = 80$ kpc. 

The length of the high energy stage may be estimated as the typical length of the leading-particle chain $L_{lead} \sim \lambda/f_{in}$; $L_{lead} \sim 300-500$ kpc for $E_{0} = 100$ PeV and  $L_{lead} \sim 3-5$ Mpc for $E_{0} = 1$ EeV. Beyond the distance of $(2-3)L_{lead}$ the cascade enters the particle multiplication stage when the energy is quickly (length scale $\sim (2-3) \lambda$) transferred to many secondaries; the number of these secondaries after the generation $N_{casc}$ is $2^{N_{casc}}$.\footnote{during the particle multiplication stage the effect of interactions on the URB as well as the TPP and DPP processes could be safely neglected}

For the range of parameters of the present study interactions on the URB as well as the TPP and DPP processes have one common feature: these are mostly near-threshold interactions with the average inelasticity $f_{in} \sim 0.5$. In this case, the particle multiplication stage sets in much faster than in the case when additional processes are neglected. The effects of the DPP process are qualitatively similar to those of the TPP process, but DPP is relevant at higher energies than TPP.

We illustrate these qualitative considerations and estimates with detailed numerical simulations performed with the CRPropa3 code in Appendix~\ref{appendixc}. In particular, we show that in the case when all three processes under investigation are accounted for, almost all of $\gamma$-rays have the energy below 200 TeV already after the propagated distance of 10 Mpc. After this distance, the ELMAG code is fully applicable.

In the present work, it will be remembered, we assume weak universality (\citep{Berezinsky2016}, see Subsect.~\ref{ssec:simulations-spectrum}); this assumption is even better justified if the interactions on the URB as well as the TPP and DPP processes are included into calculations. We have also checked that for the propagated distance comparable to the distance from the source to the observer the shape of the observable spectrum is indeed well described by the distant-dependent universal spectrum. Finally, we note that a comparison of the observable spectra in the weak universality approximation and in the ``full hybrid approach'' (without this assumption) is presented in \citet{Dzhatdoev2017}. In particular, for the redshift of the source $z > 0.02$ the weak universality assumption is valid even if one does not account the interactions on the URB as well as the TPP and DPP processes.

\subsection{Collective (plasma) energy losses \label{ssec:plasma}}

Cascade electrons may lose energy through the excitation of plasma instabilities \citep{Broderick2012}. The impact of such collective (plasma) energy losses on the observable spectrum is, at present, not well understood (e.g. \citep{Schlickeiser2012,Miniati2013,Chang2014,Sironi2014,Menzler2015,Kempf2016,Vafin2018,Vafin2019}). For this reason we did not account the plasma losses in the present work. The inclusion of this additional channel of energy losses would decrease the observable $\gamma$-ray intensity.

\section{Conclusions} \label{sec:conclusions}

In this work we calculated the observable spectral and angular distributions of very high energy cascade $\gamma$-rays as expected from extragalactic sources of ultrahigh energy protons. In particular, we accounted for the effect of primary proton deflection in an inhomogeneous extragalactic magnetic field. We demonstrated that the observable spectrum reveals a high-energy cutoff in comparison with the spectrum averaged over all values of the observable angle $\theta_{obs}$, and that the observable angular distribution is broad enough to be readily detected by the next-generation $\gamma$-ray telescopes, and, possibly, by some of the currently operating ones. To our knowledge, these results have never been reported before. We have discussed how different choice of parameters (such as source redshift, primary proton energy and angular distribution, extragalactic magnetic field and extragalactic background light models) would change the observable spectra. We have also provided a qualitative discussion of the main stages of cascade development and the effects of additional processes.

This study could significantly facilitate future axion-like particle searches in the optically thick region of blazar spectra. In addition, our work might be useful in extreme TeV blazar studies as well as in several related branches of $\gamma$-ray astronomy, including extragalactic background light and extragalactic magnetic field measurements.
  
\section*{acknowledgements}

We are greatly indebted to Dr. S. Hackstein for guiding us through the CRPropa3 simulations and to E.I. Podlesnyi for running some of these simulations while the first author of this paper was on a duty journey. We are grateful to the members of the University of Tokyo, Institute for Cosmic Ray Research (ICRR) Cherenkov cosmic $\gamma$-ray group, especially to Prof. M. Teshima. We acknowledge helpful discussions with Dr.~O.E.~Kalashev, Dr. G.I. Rubtsov, Prof. S.V. Troitsky, Dr. A.V. Uryson, as well as with some of the participants of the Extreme-2019 and TeVPA-2019 conferences. We are grateful to the anonymous referee for the timely review, and for the suggestion to consider the role of triplet pair production, double pair production, and interactions on the universal radio background. This work was supported by the Russian Science Foundation (RSF) (project no. 18-72-00083). 

\section*{Data availability statement}

The data underlying this article will be shared on reasonable request to the corresponding author.

\bibliographystyle{mnras}
\bibliography{IHCM-Broadening.bib}

\newpage

\begin{appendix}
\section{The impact of fluctuations on the observable spectra} \label{appendixa}

\begin{figure}
\centering \vspace{1pc}
\includegraphics[width=8.4cm]{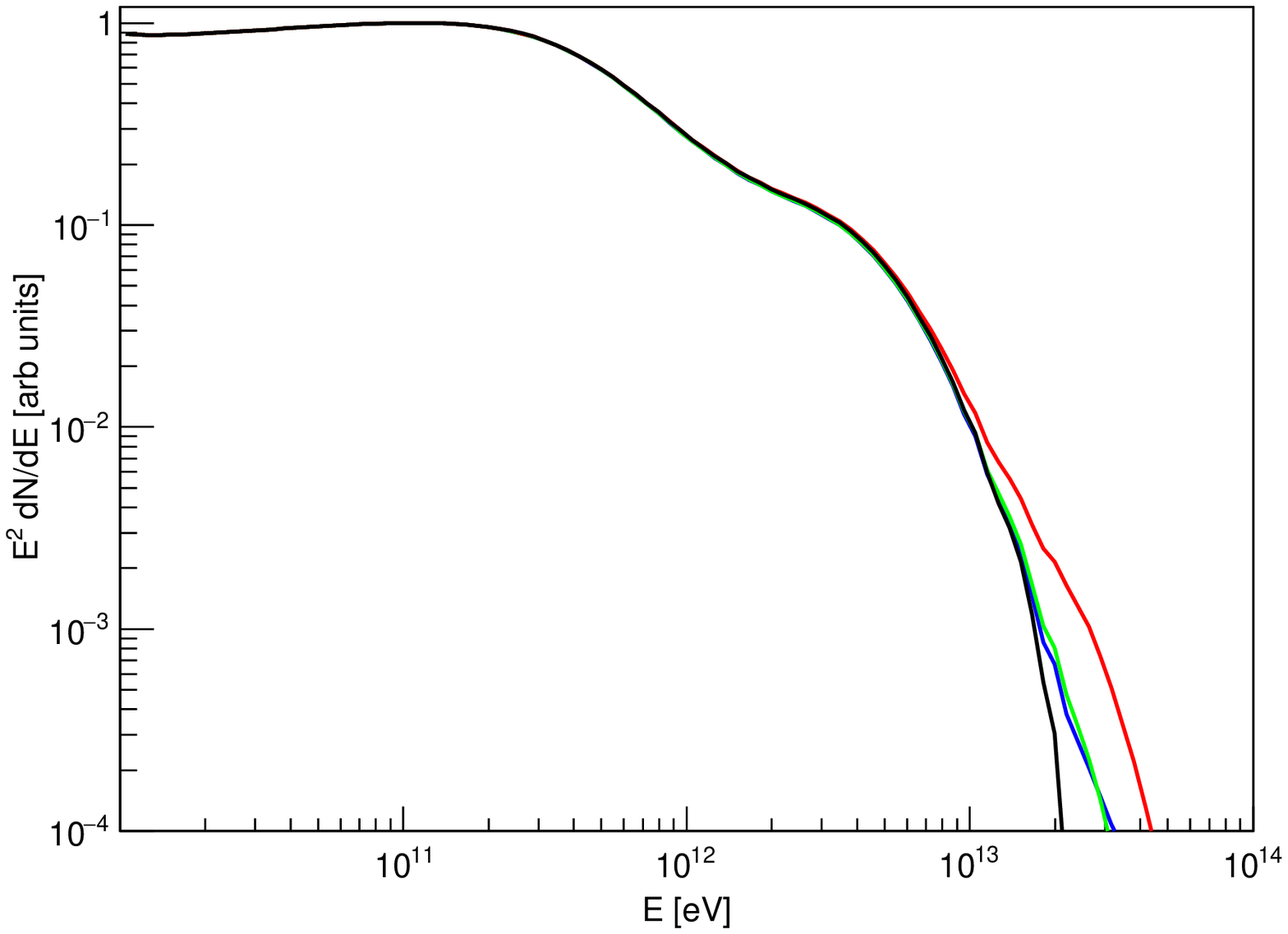}
\caption{Average observable SEDs for the KD10 EBL model option and the following numbers of trajectories: 10 (black curve), 100 (red curve), 10$^{3}$ (green curve), 10$^{4}$ (blue curve).}
\label{FigA1}
\end{figure}

The observable spectrum in the framework of the MHCM depends on the EGMF structure, in particular, on the number of proton trajectories used in our study. In Fig.~\ref{FigA1} we show the observable SEDs averaged over various numbers of proton trajectories. We conclude that the statistics of 10$^{2}$ trajectories is sufficient to obtain a stable enough average observable spectrum for $E < 10$ TeV, and the use of 10$^{3}$ trajectories already allows one to robustly determine the shape of the spectrum for $E < 30$ TeV.

\begin{figure}
\centering \vspace{1pc}
\includegraphics[width=8.4cm]{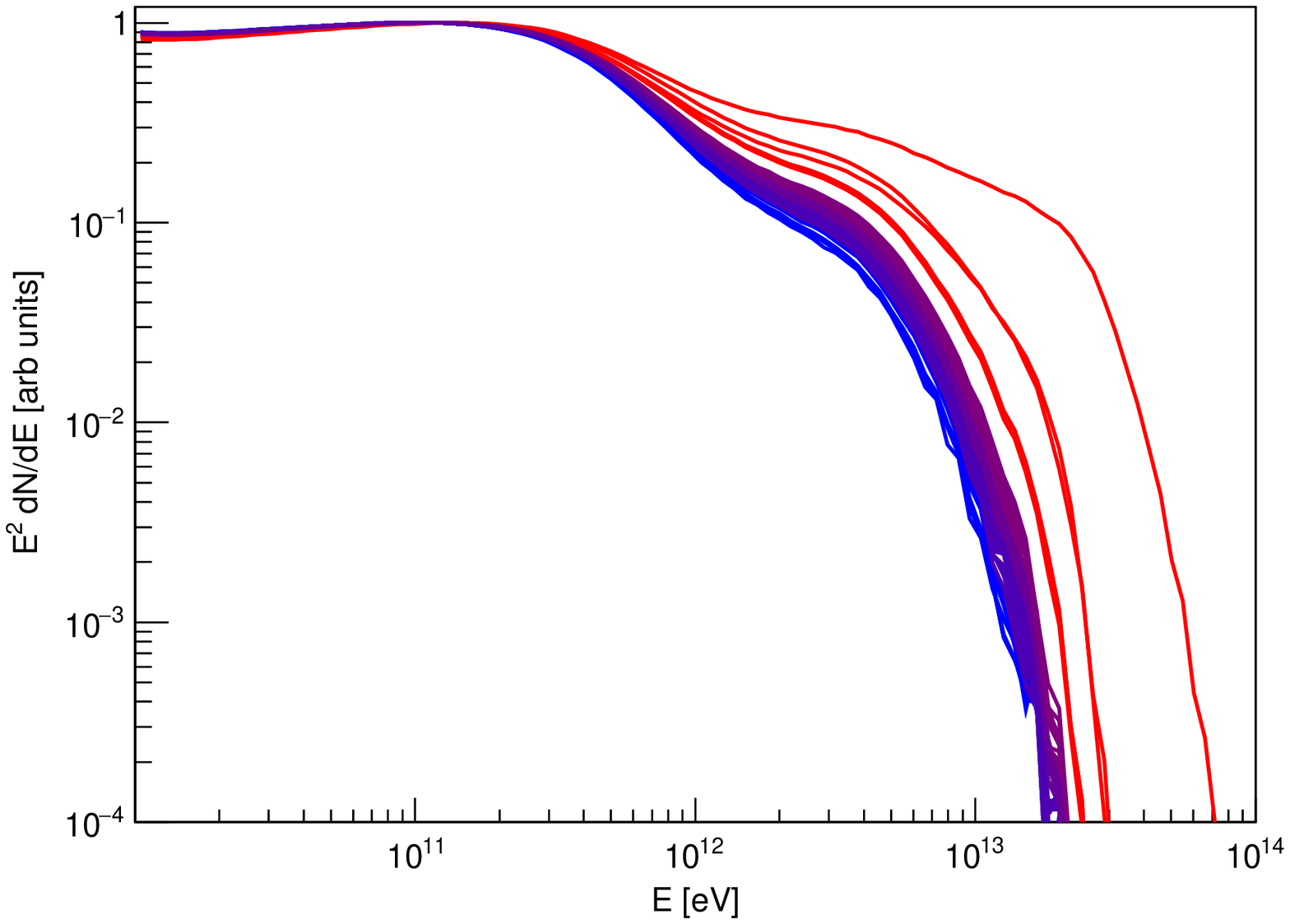}
\caption{Observable SEDs for the KD10 EBL model option and 10$^{2}$ individual trajectories; colors represent the ratios of individual SEDs at 10 TeV and 100 GeV.}
\label{FigA2}
\end{figure}

In addition, we have calculated the observable SEDs for 10$^{2}$ individual proton trajectories to demonstrate the impact of fluctuations on the shape of the observable spectrum (see Fig.~\ref{FigA2}). The blue-to-red gradient of colors gives a visual representation to the values of the ratios of different individual SEDs at two benchmark energies: 10 TeV and 100 GeV. Blue colors represent lower ratios, while red colors --- higher ones. We note that at $E = 5$ TeV the maximal difference between the flux corresponding to 95 \% of the trajectories is limited to the factor of 2.5. 

\section{Distribution of the critical distance} \label{appendixb}

\begin{figure}
\centering \vspace{1pc}
\includegraphics[width=8.4cm]{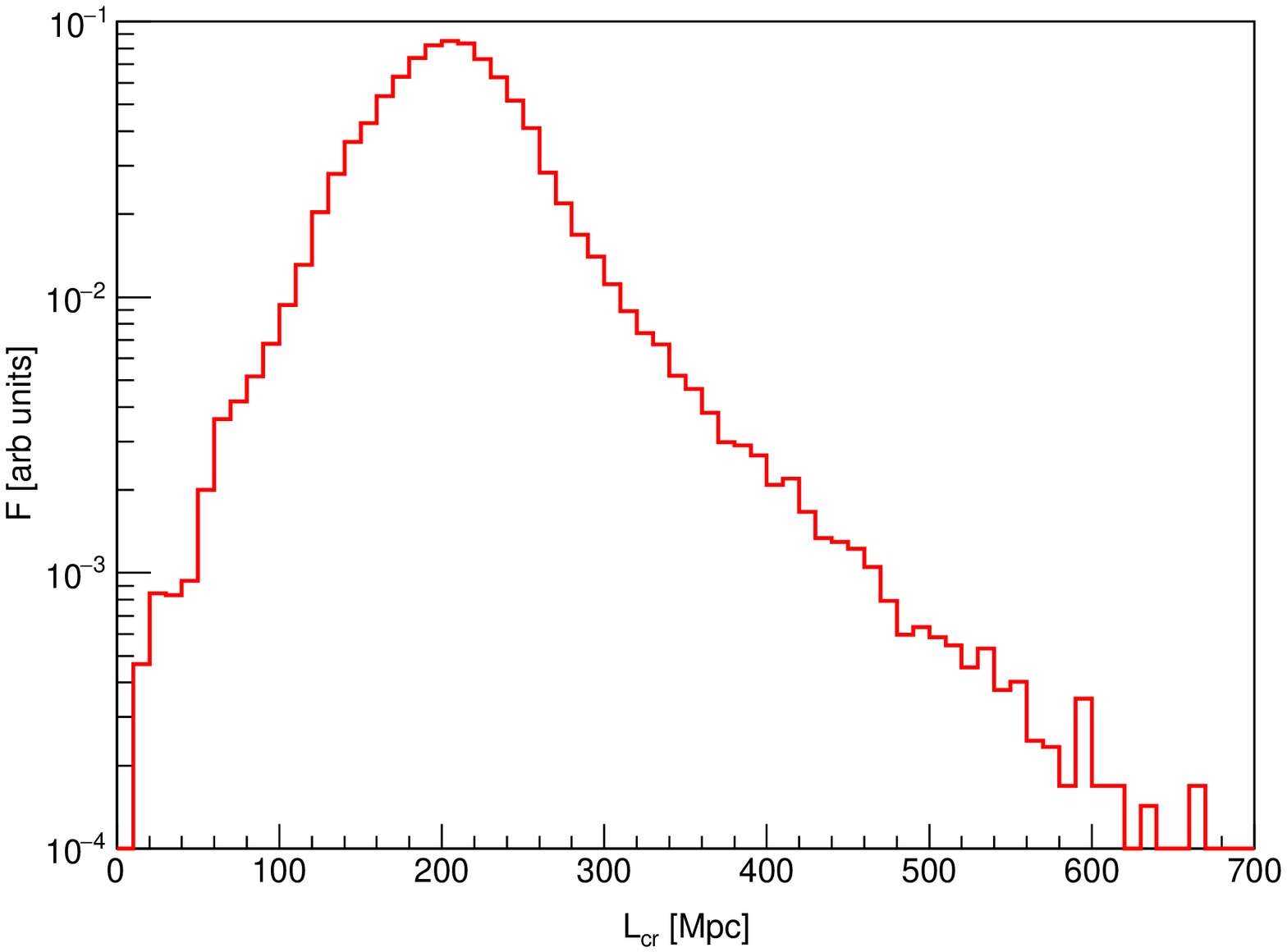}
\caption{Histogram of the critical distance $L_{cr}$ distribution (red curve).}
\label{FigB1}
\end{figure}

In Fig.~\ref{FigB1} we present a histogram of a distribution on ``the critical distance'' $L_{cr}$ (see Subsect~\ref{ssec:results-spectrum} for definition of this parameter). The normalization of this histogram corresponds to the case of discrete probability density function, i.e. $\sum_{i=0}^{N}F_{i} = 1$ with summation over the bins of the histogram, with the $N$th bin containing the largest value of $L_{cr}$.

\section{Spectra with and without TPP, URB, DPP} \label{appendixc}

\begin{figure}
\centering \vspace{1pc}
\includegraphics[width=8.4cm]{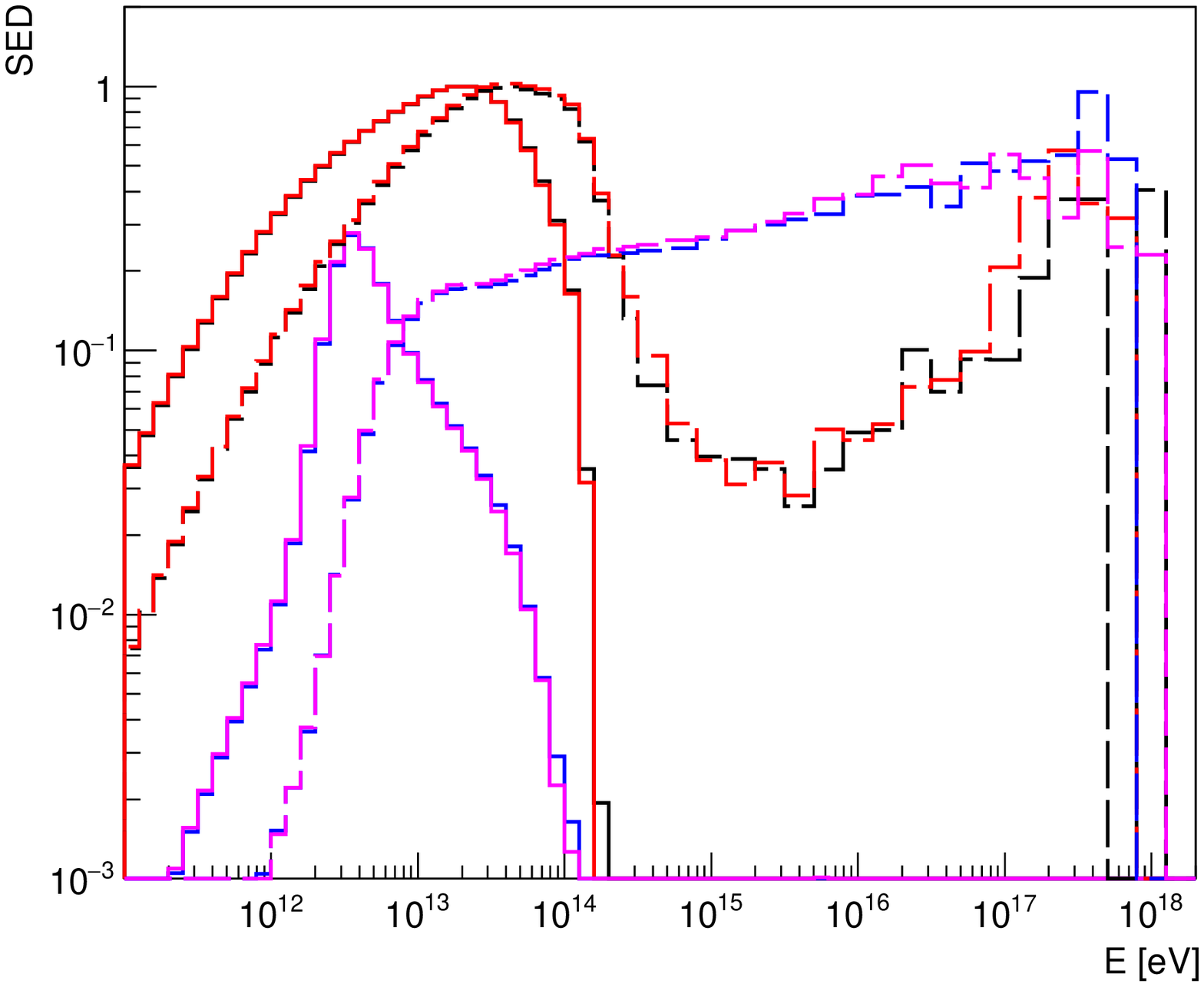}
\caption{Observable spectra of $\gamma$-rays and electrons with the account of the interactions on the URB and the TPP and DPP processes (solid curves) as well as the same spectra neglecting these processes (dashed curves). Black histograms denote observable spectra of $\gamma$-rays from primary $\gamma$-rays, red histograms --- spectra of $\gamma$-rays from primary electrons, blue histograms --- spectra of electrons from primary $\gamma$-rays, magenta histograms --- spectra of electrons from primary electrons.}
\label{FigC1}
\end{figure}

In Fig.~\ref{FigC1} we present histograms of the observable spectra of \mbox{$\gamma$-rays} and electrons calculated with the CRPropa3 code for the case of the primary $\gamma$-rays and electrons with the energy of 1 EeV and the distance between the source and the observer of 10 Mpc. $\gamma$-ray spectra obtained with the account of the interactions on the URB and the TPP and DPP processes have a marked energy cutoff at 100 TeV. Moreover, in this case the intensity of observable electrons is subdominant with respect to $\gamma$-rays. On the contrary, when the interactions on the URB and the TPP and DPP processes are neglected, observable $\gamma$-ray and electron spectra appear to be much harder.

To conclude, for the case of $E_{0} = 1$ EeV the difference between the observable spectra after the propagated distance of 10 Mpc with and without account of URB, TPP, and DPP is significant (cf. \citet{Mastichiadis1994}: ``triplet production is particularly important (...) for propagation over small path length''), but for the propagation distances much larger than 10 Mpc the observable spectra are almost the same (again, cf. \citet{Mastichiadis1994}: TPP ``has no effect on saturated cascades ''). 

\end{appendix}

\bsp	
\label{lastpage}
\end{document}